%% file: ai2ase.tex
\documentclass[letterpaper]{article} % DO NOT CHANGE THIS
\usepackage{aaai2026}  % DO NOT CHANGE THIS
\usepackage{times}  % DO NOT CHANGE THIS
\usepackage{helvet}  % DO NOT CHANGE THIS
\usepackage{courier}  % DO NOT CHANGE THIS
\usepackage[hyphens]{url}  % DO NOT CHANGE THIS
\usepackage{graphicx} % DO NOT CHANGE THIS
\usepackage{natbib}  % DO NOT CHANGE THIS AND DO NOT ADD ANY OPTIONS TO IT
\usepackage{caption} % DO NOT CHANGE THIS AND DO NOT ADD ANY OPTIONS TO IT
\usepackage{algorithm}
\usepackage{algorithmic}

\usepackage{newfloat}
\usepackage{listings}
\DeclareCaptionStyle{ruled}{labelfont=normalfont,labelsep=colon,strut=off} % DO NOT CHANGE THIS
\floatstyle{ruled}
\newfloat{listing}{tb}{lst}{}
\floatname{listing}{Listing}
\title{When Less is More: \\A Story of Failing Bayesian Optimization Due to Additional Expert Knowledge}
\author{
    Dorina Weichert\textsuperscript{\rm 1},
    Gunar Ernis\textsuperscript{\rm 1},
    Marvin Worthmann\textsuperscript{\rm 1}
    Peter Ryzko\textsuperscript{\rm 2},
    Lukas Seifert\textsuperscript{\rm 3}
}
\affiliations{
    \textsuperscript{\rm 1}Fraunhofer Institute for Intelligent Analysis and Information Systems IAIS\\
    Schloss Birlinghoven 1,
    53757 Sankt Augustin, Germany \\
    \textsuperscript{\rm 2}Südpack Verpackungen SE \& Co. KG\\
    Jägerstraße 23, 
    88416 Ochsenhausen, Germany\\
    \textsuperscript{\rm 3}Institute for Plastics Processing in Industry and Craft at RWTH Aachen University\\
    Seffenter Weg 201, 52074 Aachen, Germany\\    

}

\usepackage{bibentry}
\usepackage[acronym,numberedsection,shortcuts,nonumberlist]{glossaries}
\setacronymstyle{long-short}
\input{glossary}

\usepackage{xcolor}
\newcommand{\dw}[1]{{\color{teal}DW: #1}}
\newcommand{\ger}[1]{{\color{violet}GE: #1}}

\usepackage{subcaption}

\usepackage{siunitx}

\usepackage{comment}
\begin{document}

\maketitle

\begin{abstract}
The compounding of plastics with recycled material remains a practical challenge, as the properties of the processed material is not as easy to control as with completely new raw materials. For a data scientist, it makes sense to plan the necessary experiments in the development of new compounds using \gls{BO}, an optimization approach based on a surrogate model that is known for its data efficiency and is therefore well suited for data obtained from costly experiments. Furthermore, if historical data and expert knowledge are available, their inclusion in the surrogate model is expected to accelerate the convergence of the optimization. In this article, we describe a use case in which the addition of data and knowledge has impaired optimization. We also describe the unsuccessful methods that were used to remedy the problem before we found the reasons for the poor performance and achieved a satisfactory result. We conclude with a lesson learned: additional knowledge and data are only beneficial if they do not complicate the underlying optimization goal.
\end{abstract}

\section{Introduction}
While the EU aims for a fully circular economy by 2050 \cite{eu:2050}, only 14~\% of materials used are recycled \cite{plastics:2025}. Plastics pose particular challenges due to their extensive industrial integration and difficulty in substitution. Incorporating recycled plastics into manufacturing faces a trade-off: recycled materials have less predictable and more variable properties than virgin plastics.

This challenge intensifies when developing compounds with specified properties. Recycled materials contain unknown contaminants and degradation products that cannot be easily characterized, making property prediction very difficult and necessitating extensive experimental validation. This creates a significant bottleneck, as each experiment consumes resources, time, and skilled labor.

Traditional experimentation approaches rely on expert knowledge and carefully designed experiments developed by research and development engineers. While valuable, this expertise-driven methodology may not yield the most efficient experimental sequences, particularly for complex, non-linear relationships in recycled material compounds.

In our study, we aim to incorporate expert knowledge into the surrogate model employed by \gls{BO}, a global optimization method well-known for its sample-efficiency \cite{Garnett2023}, including applications in material science \cite{liang2021benchmarking}. The motivation for this approach stems from \gls{BO}'s proven ability to efficiently navigate complex parameter spaces while requiring fewer experimental evaluations compared to traditional methods.

However, our initial implementation of \gls{BO} yielded unexpectedly poor results, performing worse than the established \gls{DoE} methodologies developed by experienced engineers. This counterintuitive finding prompted a systematic investigation into the underlying causes. Through several iterative attempts to address the suboptimal behavior, we identified that the incorporation of expert knowledge through additional features inadvertently transformed the optimization problem into a high-dimensional space, making it more complex than necessary and compromising the efficiency that \gls{BO} is designed to provide.

Our main contributions are twofold: first, we demonstrate a practical \gls{BO} application in real-world materials development for recycled plastic compounds; second, we provide an extensive description of typical pitfalls encountered when implementing \gls{BO} in industrial materials science applications, along with a strategy to overcome them. These insights are particularly valuable for practitioners seeking to apply advanced optimization techniques in materials engineering contexts where expert domain knowledge must be carefully balanced with algorithmic efficiency.

The remainder of this paper is structured as follows: We begin with an overview of the relevant literature, followed by specifying the problem setup. Afterward, we describe the benchmarking experiments by the engineers and then move over to the failed and successful \gls{BO} approaches. We conclude by a short discussion of our experiences and an outlook how to avoid the pitfalls we have encountered.

\section{Related Work}
\paragraph{Applications of Bayesian Optimization in Materials Science}
\gls{BO} has proven effective in materials science applications due to its sample efficiency with costly experiments. \citet{liang2021benchmarking} benchmark \gls{BO} across various materials science use cases, establishing its viability for materials discovery. \citet{cinquin2025what} explore \gls{BO} integration with \gls{LLM} for high-dimensional chemistry problems, finding that simple, well-initialized surrogate models with feature fine-tuning often outperform complex approaches.
\paragraph{Practical Frameworks and Implementation Challenges}
Several frameworks facilitate practical \gls{BO} implementation, including Ax \cite{olson2025ax} and BayBE \cite{BayBE2025}, addressing multi-objective optimization and constraint handling in industrial settings. However, practical applications can encounter significant challenges. \citet{Rijn_2025} identify failure modes in \gls{BO}, specifically boundary oversampling issues where algorithms disproportionately sample parameter space boundaries, leading to suboptimal exploration.
\paragraph{Predictive Modeling of Polymer Compounds}
\citet{seifert2024a,seifert2024b} developed mechanistic models and symbolic regression approaches for binary and ternary polypropylene compounds, predicting \gls{MFR} and Young's modulus—key objectives in our study. However, their work focuses on simplified systems without additives, which are standard in industrial applications and significantly influence material properties.
\paragraph{Research Gap}
The literature reveals a gap between theoretical \gls{BO} capabilities and practical industrial implementation. Limited attention has been given to systematic identification and resolution of implementation pitfalls in real-world materials development. Our work addresses this gap through a case study of \gls{BO} implementation in industrial recycled plastic compound development.

\section{Problem Setup}
In our study, four different raw materials are mixed to generate a new plastic compound: a virgin polypropylene, recycled plastics from a local plastics recycling company, a filler material, and a so-called impact modifier that changes specific properties of the compound. 
After processing a recipe of these four ingredients, the quality of the compound is assessed using three different quality metrics: the \gls{MFR}~\cite{ISO1133,ASTMD1238} that describes the viscosity of the raw material at processing temperatures, the viscosity being a useful measure for the processability, the Young's modulus~\cite{ISO527,ASTMD638} that describes the plasticity of the final product, and the impact strength~\cite{ISO179,ASTMD256} that describes how tough and resilient a product is against applied mechanic forces.

For these quality metrics, the engineers defined the following objective values: a \gls{MFR}-value close to \SI{10}{g/ 10 min}, a Young's modulus of at least \SI{1500}{MPa}, and an impact strength of at least \SI{8}{kJ/m^2}. These values should result in good processability of the compound and good product properties for everyday use. More specifically, the resulting compound is to be used in the manufacture of thin plastic bags for foodstuffs such as chips or sweets. 

For the ingredients, the engineers defined the following bounds: while the proportion of the virgin propylene and recycled plastics in the mixture can be up to 100~\%, the filler is limited to 30~\%, and the impact modifier to 20~\%. 

Given these key figures, we formulate a constrained optimization problem: minimize the difference to the objective \gls{MFR}-value, with the Young's modulus and the impact strength being above their given limits (i.e., output constraints), and the values of the inputs being in their limits and summing up to one (because we design a mixture).

Provided with this problem definition, we can start the experimentation process.

\section{Experimental Baseline}
\begin{figure}
    \centering
    \includegraphics[width=\columnwidth]{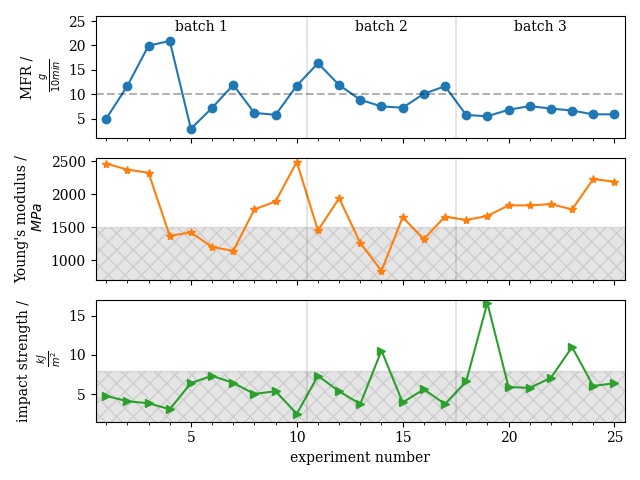}
    \caption{Values of quality metrics based on experiments by engineers.}
    \label{fig:performanceengineer}
\end{figure}%
The experimental campaign conducted by process engineers followed a batched approach, comprising 25 experiments organized into three sequential batches: an initial batch of ten compounds, followed by seven compounds, and concluding with eight compounds. This batched experimentation strategy was necessitated by the costly and time-intensive evaluation of quality metrics. While \gls{MFR} measurements can be performed immediately after compound preparation, the assessment of Young's modulus and impact strength requires the production of samples with defined geometry from each compound, followed by standardized mechanical testing procedures. This multi-step process significantly increases both time requirements and experimental costs.

The experimental timeline spanned two days: the first batch was completed on day one, while batches two and three were conducted on day two. The performance results of these engineer-designed experiments are presented in figure~\ref{fig:performanceengineer}.

A significant challenge emerged during the experimental campaign: identifying parameter combinations that yielded sufficient impact strength while maintaining acceptable levels of the Young's modulus. 
The data partially revealed complex ingredient interactions responsible for the opposing behavior observed between Young's modulus and impact strength in the first two batches. Specifically, experiments demonstrated an inverse relationship between these properties: when impact strength was high (as observed in experiment 14), Young's modulus tended to be low, and vice versa (as demonstrated in experiment 10). This trade-off relationship highlights the inherent complexity of recycled plastic compound formulation.

Despite these challenges, the final batch yielded promising results, with two experiments fulfilling all specified constraints. The best reached value for the \gls{MFR} is \SI{6.65}{g/10 min}. This outcome validated the engineers' iterative approach and demonstrated the feasibility of achieving the desired property combination through systematic experimentation, albeit at considerable time and resource investment.

\section{Performing \acrlong{BO}}
The theoretical advantages of \gls{BO} - particularly its sample efficiency and ability to balance exploration and exploitation - make it a promising candidate for optimizing expensive experimental campaigns in materials development. Therefore, we expected \gls{BO} to reveal comparable or superior results to the state-of-the-art \gls{DoE} methodologies employed by experienced engineers.

To ensure a fair comparison with the expert-conducted experiments, we designed our \gls{BO} implementation to generate batches of identical sizes to those used in the real-world expert \gls{DoE}: ten experiments in the first batch, followed by seven in the second batch, and eight in the third batch, totaling 25 experiments. This batched approach mirrors the practical constraints faced by the engineering team, where experimental evaluations must be conducted in groups due to equipment availability, processing time, and resource allocation considerations.

Our \gls{BO} implementation strategy encompasses several approaches, beginning with sophisticated models that incorporate expert knowledge through additional features, and progressively simplifying the problem formulation based on observed performance. This iterative refinement process provides valuable insights into the practical challenges of deploying \gls{BO} in industrial materials development contexts.

\subsection{Modeling the Compound Properties Using Expert Information}
The performance of \gls{BO} relies on the efficient handling of data via a probabilistic model \cite{Garnett2023}. Expert information can improve the model and make the approach even more sample-efficient, which is what we intended to do via the following approach:

The engineers provided us with a set of 430 experiments performed with different compositions of virgin and recycled plastics, impact modifiers, filler materials and additives at different production parameterizations.
From these experiments, we considered experiments with nine types of virgin plastics, three types of recycled plastics, a single impact modifier, and three types of fillers, as, for these materials, we were also given data sheets. 
Furthermore, we filtered the experiments for the production parameters (the configuration of the screw of the extruder, the temperature in the extruder's end zone, the feed rate, and the rotation speed) matching the parameter configuration used by the engineers.

Given the data sheets, we generated features for a generic model of the behavior of the compound. Therefore, for each of the main components (virgin and recycled plastics, impact modifier, and filler), we determined their proportion and their expected impact on the quality metrics via the data sheet, and we find an eleven-dimensional problem. After data cleaning, we were left with a dataset of 50 instances to train a \gls{GP} regression model in the state-of-the-art botorch framework~\cite{balandat2020botorch} - a rather rough model, but it is only used provide the \gls{BO} approach with a general idea of the material behavior.

\begin{figure}
    \centering
    \includegraphics[width=0.95\linewidth]{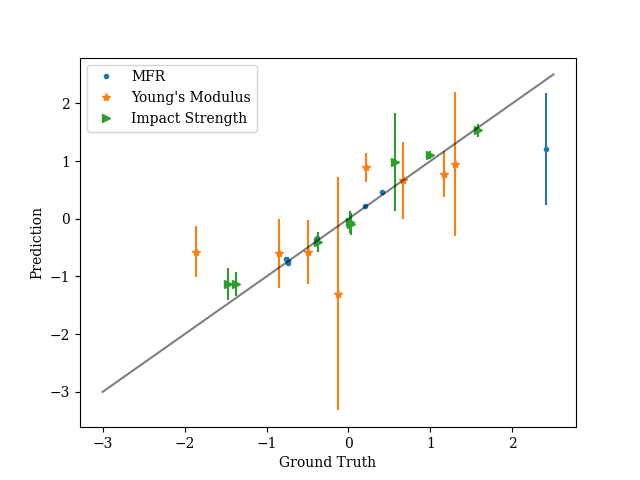}
    \caption{Predictive performance of \gls{GP} model on test data set, the root-means-squared-errors given in scaled space.}
    \label{fig:performanceGP}
\end{figure}

We train the model making a train-test-split with 85~\% train data and assess its performance using the test data. 
For the model, we used a multioutput \gls{GP} with uncorrelated outputs, but trained on the sum of the marginal log likelihoods. For each model, we use a squared-exponential covariance function and a zero mean function, which are defaults in the used framework. We scale the features of our data to a \(\left[0, 1\right]^{11}\) bounding box according to the input limits provided by the experts and use a standard scaling for the outputs of the model. 

In figure~\ref{fig:performanceGP}, we provide the quality of the model's predictions in scaled space. We can see that the model is in most cases very confident where its predictions meet the ground truth and unconfident far away from it. Overall, it is not very exact, but it covers the data sufficiently.

Before performing \gls{BO}, we add all data to the model, while fixing its hyperparameters.

\subsection{Our Initial Oracle}
To assess the quality of experiments proposed by \gls{BO}, we developed a predictive model joining the cleaned historical dataset with the results from the engineer-conducted experiments.

We implemented \gls{GP} regression models for the prediction of each compound quality measure (\gls{MFR}, Young's modulus, and impact strength), employing the defaults for \gls{GP}es with a fully Bayesian treatment of the hyperparameters in \citet{balandat2020botorch}, building on the results of \citet{pmlr-v235-hvarfner24a}.

The predictive quality of each model was rigorously assessed via leave-one-out cross-validation, which provides an estimate of model performance on unseen data. This validation approach is particularly appropriate for our limited dataset size, as it maximizes the use of available training data while providing reliable performance estimates. The cross-validation yielded an average root means squared error of \SI{2.23}{g/10 min} for the \gls{MFR}, of \SI{2.04}{kJ/m^2} for the impact strength, and of \SI{152}{MPa} for the Young's modulus, indicating a good prediction quality. 

The predictive mean of these trained models serves as an oracle in subsequent \gls{BO} experimentation, enabling systematic evaluation of proposed experimental designs without requiring costly physical experiments. This oracle-based approach allows us to assess the efficiency and effectiveness of different \gls{BO} strategies while maintaining experimental realism through models trained on actual compound data.

\subsection{Failed Runs of \glsentrylong{BO}}
We implement the problem setup into the \gls{BO} procedure. To reach the specified target value of the \gls{MFR}, our main optimization objective is to minimize the quadratic distance to it. For the constraints, we multiply the acquisition function with their probability of feasibility, following the traditional approach by~\citet{Gardner2014}. 

As acquisition function, we employ the log noisy expected improvement acquisition function, which is expected to be robust against the potential output noise \citep{Ament2023}.

As the experimentation is performed using a fixed set of raw materials with specified properties available in data sheets, we include this data and fix the corresponding feature values during the optimization of the acquisition function. Also during acquisition function optimization, we add the mixture constraint.

The resulting runs of \gls{BO} that we present in the following were created using a single random seed, as in real life, we would also make only one run and not a higher number of repetitions as we usually apply in technical papers. Nevertheless, we ran the experiments multiple times to ensure that the results are representative for the course of experimentation.

\subsubsection{Run 1: Vanilla Constrained \glsentrylong{BO}}
The first run that uses our model with expert knowledge failed totally. The acquisition function optimizer was not able to find input regions where the probability of feasibility for both the constraints on the impact strength and on the Young's modulus is above zero simultaneously. An inspection of the data used to train the model revealed that there is no historical data where both constraints are fulfilled simultaneously and that there are only two instances with the impact factor fulfilling the constraint, thus, it is highly probable that this is the reason for the bad behavior.

\subsubsection{Run 2: Iterative Relaxing of Constraints}
\begin{figure}
    \centering
    \includegraphics[width=\columnwidth]{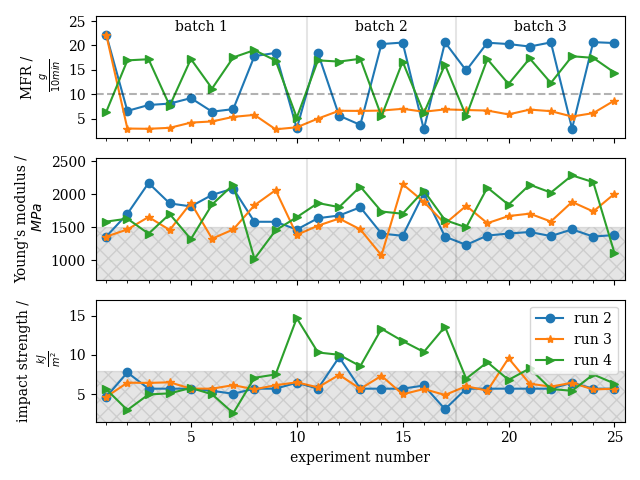}
    \caption{Course of experimentation for \gls{BO} approaches. Run 4, with the reduced model is leads to more proposed experiments matching the constraints than in runs 2 and 3, as well as than the manual experimentation.}
    \label{fig:performanceBO}
\end{figure}%
As we found that the constraints are a problem, we tried to relax them for the first and second batch of experiments. Therefore, we implemented a gradual reduction of the respective constraints on the Young's modulus and the impact strength until the acquisition function optimization was successful and provided a batch of experiments. Unfortunately, the results of this procedure, see figure~\ref{fig:performanceBO}, do not meet our expectations as the constraint on the impact strength is met only once, associated with a \gls{MFR} of \SI{5.60}{g/10 min}.

\subsubsection{Run 3: Reformulating the Optimization Problem}
Afterward, we tried to use \gls{BO} to generate data with high values of the impact strength, as these seem to be underrepresented in the data. Therefore, for the first two batches, we tried to find regions where this constraint is met, reformulating the optimization problem such as to maximize the impact strength. Only in the last batch, we apply the known constraint on the Young's modulus and add a constraint on the \gls{MFR} such as to be inside a corridor of \(10 \pm\)\SI{5}{g/10 min}. We provide the results in figure~\ref{fig:performanceBO}. Again, they do not meet the expectations, as the constraint on the impact strength is met only once, associated with a \gls{MFR} of \SI{5.90}{g/10 min}.

\subsection{The Working, Simple Approach}
\paragraph{Challenges with High-Dimensional Modeling and Oracle Quality}
The results obtained for the reformulated optimization problem raised significant suspicions regarding the quality of our oracle model and the validity of our overall approach. Upon closer examination, we identified several critical issues that compromised the effectiveness of our \gls{BO} implementation.

A primary concern emerged from the limited amount of available data: our oracle was trained on only 75 data points spanning 11 dimensions, raising fundamental questions about the curse of dimensionality~\citet{bellman1957}. With such sparse coverage of the input space, it became questionable whether sufficient data existed to train a reliable predictive model capable of accurate interpolation across the entire search space.

This data sparsity issue was further compounded by the well-known boundary issue in \gls{BO} \cite{swersky2017improving}, which becomes particularly problematic in medium to high-dimensional spaces. The boundary issue manifests as a tendency for \gls{BO} algorithms to disproportionately explore the boundaries of the search space, driven by the larger predictive variance of \gls{GP} models at boundaries compared to the interior volume where interpolation can be used to estimate variance more accurately.

While extensive literature exists on \gls{BO} approaches for high-dimensional problems—including subspace methods, additive structure assumptions, local optimization techniques, and non-Euclidean covariance functions, as comprehensively reviewed by \citet{pmlr-v235-hvarfner24a}—these sophisticated methods proved unsuitable for our specific case. The fundamental limitation was our severely constrained experimental budget of only 25 to at most 75 experiments, which is substantially smaller than even the typical initialization set sizes employed in high-dimensional \gls{BO} studies \cite{pmlr-v235-hvarfner24a}.

Moreover, our oracle model, trained using the methodology described in \citep{pmlr-v235-hvarfner24a}, exhibited a critical limitation: it lacked the ability to predict higher values of impact strength that were not in the direct neighborhood of the three instances observed during expert experimentation. This extrapolation weakness severely limited the oracle's utility for guiding optimization toward potentially superior regions of the parameter space.

\paragraph{Model Simplification and Reformulation}
To address these fundamental issues, we adopted a pragmatic approach: simplifying our models by eliminating additional features derived from technical data sheets and retaining only the four essential composition parameters. This dimensionality reduction strategy aimed to improve model reliability by focusing on the most critical input variables while mitigating the curse of dimensionality.

Correspondingly, we reformulated our oracle to utilize exclusively the 25 instances from real-life experimentation, eliminating the expanded dataset that had introduced additional uncertainty. This approach prioritizes model reliability over coverage, ensuring that predictions remain grounded in actual experimental observations rather than potentially unreliable extrapolations.

The simplified oracle achieved a leave-one-out cross-validation error (as root-mean-squared-error in real space) of \SI{4.13}{g/10 min} for the \gls{MFR}, of \SI{215}{MPa} for the Young's modulus and of \SI{2.35}{kJ/m^2} for the impact strength. These scores are larger than the ones for the prior model with all features when revisiting the scores of the first model (MFR: \SI{2.23}{g/10 min}, Young's modulus: \SI{152}{MPa}, impact strength: \SI{2.04}{kJ/m^2}), but, as the considered sample size is a third of the original sample size, it is no wonder that leaving out a single instance has more impact. Furthermore, from the comparison of the scores, we expect the model of the impact strength to have improved, as its performance has only slightly degraded.

The optimized approach with dropped features forms the foundation for our subsequent \gls{BO} investigation, providing a more reliable basis for optimization while acknowledging the inherent limitations imposed by our constrained experimental dataset.

\paragraph{Run 4: Performing \glsentrylong{BO} in the Simplified Space}
To finally benchmark \gls{BO} against the expert-based \gls{DoE}, we take the following course of experimentation: The first batch of ten experiments is made up of random data gained by rejection sampling from a Dirichlet distribution. Samples that are rejected, if they do not meet the bounds of the input domain.

After evaluating the oracle, a surrogate model is trained via maximizing the marginal log likelihood and the constrained \gls{BO} procedure (use the noisy log expected improvement acquisition function to minimize distance to \gls{MFR}-target value with impact strength and Young's modulus above their given limits) is used to obtain the second batch of experiments.

After their evaluation, the model is retrained and the third batch of experiments is generated and evaluated.

\subsubsection{Results}
As can be seen in figure~\ref{fig:performanceBO}, our final approach finally reveals the expected result - \gls{BO} is indeed able to outperfom the expert-based design, as it finds more and better points that match the constraints and simultaneously minimize the distance to the targeted \gls{MFR} - overall, there are 10 runs that meet the constraints, and the best reached \gls{MFR} is \SI{6.13}{g/10 min}, a value that is close to the result by the experienced engineers.

\section{Discussion, Conclusion and Outlook}
Our study reveals that performing \gls{BO} in real-world industrial applications remains a significant challenge, despite its theoretical advantages and proven success in controlled environments. In our use case, we attempted to leverage available expert knowledge by incorporating additional features derived from technical specifications and process parameters into our optimization framework. However, this approach ultimately failed due to the high dimensionality introduced by these features, where both the curse of dimensionality \cite{bellman1957} and the well-documented boundary issues in high-dimensional \gls{BO} \cite{swersky2017improving} hindered successful optimization.

The simplification of our problem formulation—reducing the feature space to essential composition parameters—finally led to the desired results, achieving performance comparable to expert-designed experiments. However, this outcome raises a critical question for practitioners: when and how should expert knowledge be incorporated into \gls{BO} frameworks without compromising algorithmic performance? Our experience demonstrates that the intuitive approach of adding domain-specific features can paradoxically harm optimization effectiveness, highlighting the delicate balance between incorporating expertise and maintaining computational tractability.

Future work should therefore concentrate on developing systematic approaches for assessing the suitability of \gls{BO} applications in specific industrial contexts. This could include the development of structured questionnaires or diagnostic tools to guide practitioners in problem formulation and feature selection. Additionally, the materials science and optimization communities would benefit immensely from more publications focusing on the practical dos and don'ts of \gls{BO} implementation, moving beyond theoretical contributions to address the real-world challenges encountered when deploying these powerful optimization techniques in industrial settings. Such practical guidance would accelerate the adoption of \gls{BO} in materials development and help practitioners avoid the pitfalls we encountered in this study.

\bibliography{aaai2026}
\end{document}

%% file: glossary.tex
\newacronym{BO}{BO}{Bayesian Optimisation}
\newacronym[longplural={Gaussian Processes}]{GP}{GP}{Gaussian Process}
\newacronym{DoE}{DoE}{Design of Experiments}
\newacronym{EU}{EU}{European Union}
\newacronym{RD}{R\&D}{research and development}
\newacronym{MFR}{MFR}{melt flow rate}
\newacronym{DFT}{DFT}{Density Functional Theory}
\newacronym{PP}{PP}{polypropylene}
\newacronym{LLM}{LLM}{Large Language Model}